# Data-Driven Analysis and Common Proper Orthogonal Decomposition (CPOD)-Based Spatio-Temporal Emulator for Design Exploration

Shiang-Ting Yeh[a,*], Xingjian Wang[a,**], Chih-Li Sung[b,†],
Simon Mak[b,†], Yu-Hung Chang[a,‡], Liwei Zhang,[a,##] C. F. Jeff Wu[b,§], Vigor Yang[a,¶]

[a]School of Aerospace Engineering, Georgia Institute of Technology, Atlanta, Georgia, USA
[b]School of Industrial and Systems Engineering, Georgia Institute of Technology, Atlanta, Georgia, USA

**Abstract**

The present study proposes a data-driven framework trained with high-fidelity simulation results to facilitate decision making for combustor designs. At its core is a surrogate model employing a machine-learning technique called kriging, which is combined with data-driven basis functions to extract and model the underlying coherent structures. This emulation framework encompasses key design parameter sensitivity analysis, physics-guided classification of design parameter sets, and flow evolution modeling for efficient design survey. To better inform the model of quantifiable physical knowledge, a sensitivity analysis using Sobol' indices and a decision tree are incorporated into the framework. This information improves the surrogate model training process, which employs basis functions as regression functions over the design space for the kriging model. The novelty of the proposed approach is the construction of the model through Common Proper Orthogonal Decomposition, which allows for data-reduction and extraction of common coherent structures. The accuracy of prediction of mean flow features for new swirl injector designs is assessed and the dynamic flowfield is captured in the form of power spectrum densities. This data-driven framework also demonstrates the uncertainty quantification of predictions, providing a metric for model fit. The significantly reduced computation time required for evaluating new design points enables efficient survey of the design space.

[*]Co-first author, Graduate Student, School of Aerospace Engineering
[**]Co-first author, Research Engineer, School of Aerospace Engineering
[†]Graduate Student, School of Industrial and Systems Engineering
[‡]Graduate Student, School of Aerospace Engineering
[‡‡] Research Engineer, School of Aerospace Engineering
[§]Professor and Coca-Cola Chair in Engineering Statistics, School of Industrial and Systems Engineering
[¶]Corresponding author, William R. T. Oakes Professor and Chair, School of Aerospace Engineering, vigor.yang@aerospace.gatech.edu

**Introduction**

For high-performance power generation and propulsion systems, such as those of airbreathing and rocket engines, physical experiments are extremely expensive due to the harsh requirements of operating conditions and high level of system complexities [1-3]. In addition, it is difficult to gain insight into the underlying mechanisms of the physiochemical processes involved because of the typical reliance upon optical diagnostics for experimental measurements [4-6].

High-fidelity simulations can be employed to capture more salient features of the flow and combustion dynamics in engines [7-9]. These computations, however, are often too expensive and time-consuming for design survey purposes. In the development stage for a propulsion engine, achieving an optimal design requires models capable of evaluating designs and identifying trade-offs in a timely manner. Furthermore, the formulation of such models requires an understanding of the key physics and incorporation of physics-guided decision-making to resolve multiple, potentially conflicting, requirements. The present study, for example, treats a simplex swirl injector, a central component of many airbreathing and rocket combustion devices containing rich flowfield physics. Each high-fidelity calculation of the three-dimensional flow evolution using the Large Eddy Simulation (LES) technique takes about 500,000 CPU hours to obtain statistically meaningful data for a grid of 4 million mesh points [9]. Given the number of geometric attributes and operating conditions to be surveyed, the design space exploration necessitates a prohibitive number of sample points. The situation can be substantially improved by using Design of Experiments (DoE) [10], statistical methodologies to determine the ideal training dataset for surrogate modeling. By identifying sensitive injector parameters, the sample size can be further reduced based on semi-empirical approaches, such as the recommendation of ten simulations per parameter proposed by Loeppky et al. [11].

Kriging, a technique originating in the field of geostatistics [12], is a powerful machine-learning tool for interpolation and prediction. The key idea is to model unobserved responses using a Gaussian Process (GP) governed by a preset covariance function. The response surface of the trained kriging model can then be obtained by applying data-tuned weights to radial basis functions centered at observed points. For problems of modeling spatio-temporal flow evolution, the observed points over the entire design space are sparse because the daunting computational costs limit the number of affordable numerical cases. Conventional machine-learning techniques relying on "big data" over the design space would fail. Rather, the "big data" lies within the flowfield information, which encompasses a wide range of length and time scales. The proposed methodology combines machine-learning techniques with domain knowledge of the physical system to build an accurate emulator model. The inclusion of flow physics allows the data-driven model to be physically interpretable with enhanced emulation performance.

The objective of the present study is to develop a kriging model capable of treating different spatial grids while capturing dynamic information. The spatial and temporal resolution of all simulation cases is very fine, making direct use of the raw data for training a predictor computationally demanding. Not only is there a need to incorporate data generated from different spatial grids [13-15] and use data-reduction methods, but the kriging model must also be extended to multiple, functional outputs. A limited number of studies have been published on multiple-output kriging [16, 17] and functional outputs, including wavelet decomposition [18] and knot-based GP models [19]. For data with fine spatio-temporal resolution, unfortunately, these types of methods are inappropriate because of the substantially increased computation time required.

Here, an emulation framework for a spatio-temporal surrogate model is presented using a simplex swirl injector for demonstration. A reduced basis model is developed and implemented to handle large scale spatio-temporal datasets with practical turn-around times for design iteration. The model is trained with datasets that have been classified based on established physics to reap the benefits of incorporating machine learning techniques into the framework. The model can accurately retain the rich set of physics from LES-based high-fidelity simulations and predict flow structures.

The present work develops an integrated framework that incorporates state-of-the-art statistical methods, machine learning algorithms, and a physics-driven data reduction method to obtain a surrogate model over a broad range of design space. The emulation framework relies on Proper Orthogonal Decomposition (POD) [20] (also known as the Karhunen-Loeve decomposition in the theory of stochastic processes [21]) to extract the flow physics and reduce the data by representing the flowfield with basis functions. This technique can be combined with kriging to build an efficient and physics-driven emulator. The basis Common POD (CPOD) analysis is conducted by means of a common grid generated from simulations of the geometries designated by a DoE. Such an approach is similar to that of Higdon et al. [22] for generalizing a POD expansion. In a complementary statistical paper [23], the statistical properties of a broader class of CPOD-based emulators are considered.

The present paper applies new machine-learning techniques and investigates the practical performance of the emulator with respect to flow physics. The emulated flowfield is validated against an LES simulated flowfield to demonstrate how the flow structures and injector characteristics are captured by the model. In addition, the model allows for spatio-temporal uncertainty quantification (UQ). This metric can be used to verify the model and quantify the tolerance range for flow oscillations.

The paper is structured as follows. Section 1 provides the physical model, describing the baseline configurations, the design points designated by the DoE, the high-fidelity simulation technique, and the simulation results. Section 2 discusses the data-driven emulation framework proposed for the design methodology and surrogate model.

Section 3 details the application of the framework, while assessing the surrogate model using performance metrics, root-mean-square errors, and power spectrum density (PSD) of simulated and predicted flowfields. Finally, Section 4 concludes with a summary and directions for future work.

1. **Physical Model Description and Simulations**

*1.1. Swirl Injectors*

Figure 1 shows a schematic of a simplex swirl injector representative of those commonly used in applications like liquid-fueled propulsion engines [24]. The five parameters that define the geometry are injector length, $L$, injector radius, $R_n$, inlet slot width, $\delta$, tangential inlet angle, $\theta$, and the distance between the inlet and headend, $\Delta L$. These design parameters play an important role in determining the injector performance, including the thickness, $h$, and spreading angle, $\alpha$, of the liquid film at the injector exit. The selection of these design parameters is dependent upon engine requirements. Table 1 shows the design space and the ranges of each parameter in the present work. To generalize the emulator framework, a broad span of these parameters was chosen; the injector length range represents that of small upper-stage and large first-stage engines, or about 22.7 and 93 mm for the RD-0110 and RD-170 engine, respectively [25].

Liquid oxygen (LOX) at a temperature of 120 K is delivered tangentially into the injector through inlets. The operating pressure is 100 atm, typical of contemporary liquid rocket engines. The ambient gas is oxygen at 300 K. The flow dynamics of this class of injectors have been systematically investigated in detail by Wang et al. [8] and Zong et al. [26]. Here we first conduct a set of high-fidelity simulations based on conditions in the prescribed design space described in Table 1, then extract the common flow structures for surrogate modeling.

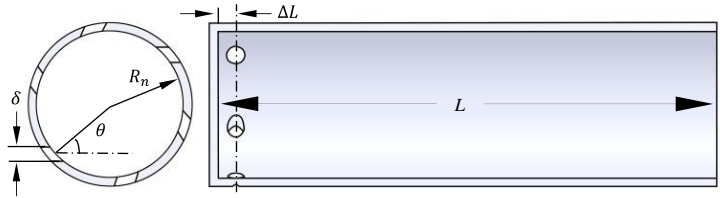

Figure 1. Schematic of a swirl injector.

Table 1 Design space for injector geometric parameters.

| $L$ (mm) | $R_n$ (mm) | $\theta$ | $\delta$ (mm) | $\Delta L$ (mm) |
|---|---|---|---|---|
| 20-100 | 2.0-5.0 | 45°-75° | 0.5-2.0 | 1.0-4.0 |

## 1.2. Design of Experiments

Given the design space in Table 1, if 10 variations are assigned for each design parameter, the total number of design points are $10^5$ for a full factorial design. It is impractical to perform so many simulations, due to the extensive computing resource required to acquire usable data. A DoE methodology is therefore required to reduce the number of design points and still capture the prominent features in the design space. To this end, the maximum projection (MaxPro) design proposed by Joseph et al. [27] is implemented for good space-filling properties and GP modeling predictions. Thirty points in the expected range of 5-10$p$ ($p$=5, the number of design parameters) points, as suggested by Santner et al. [10], are simulated over the entire design space. Figure 2 shows the two-dimensional projection of the 30 design points (and thus 30 simulation runs). Good space-filling properties are observed for all parameters.

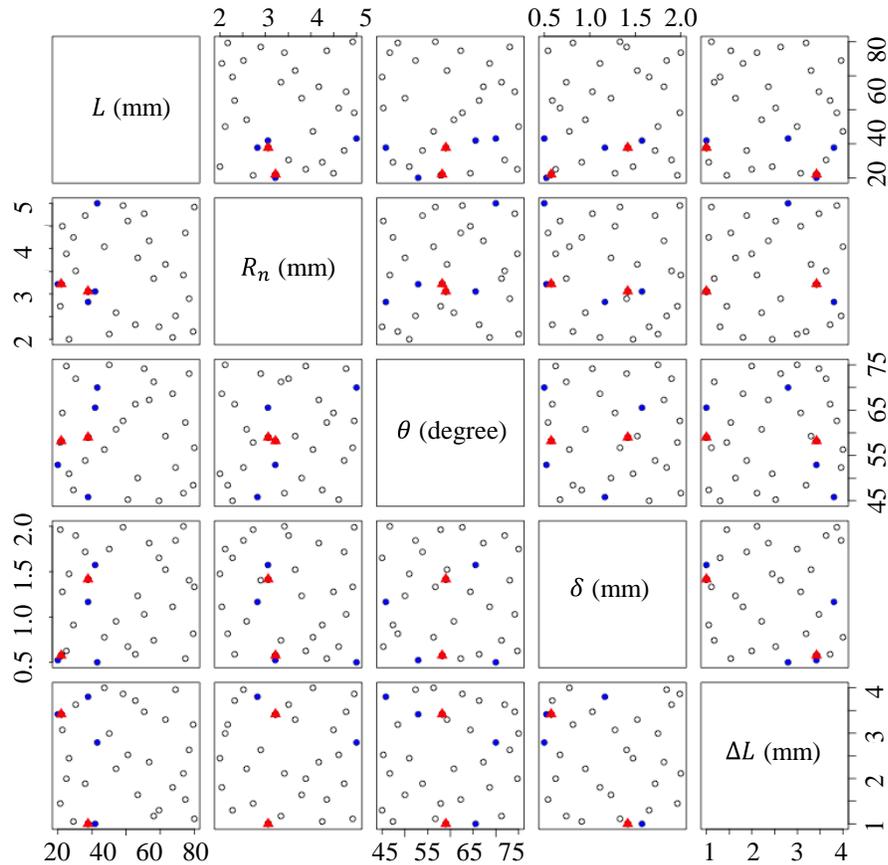

Figure 2. Two-dimensional projections of design points: benchmark points (red), baseline and neighboring points (blue).

### *1.3. High-fidelity Simulation*

An integrated theoretical and numerical framework is established and adopted, to treat supercritical fluid flows and combustion over a broad range of fluid thermodynamic states [8, 28-30]. Turbulence closure is achieved using LES.

Thermodynamic properties are evaluated by fundamental thermodynamic theories in accordance with the modified Soave–Redlich–Kwong equation of state. Transport properties are estimated using extended corresponding-state principles [28]. The numerical scheme is a density-based, finite-volume methodology along with a dual-time-step integration technique. The overall algorithm is self-consistent and robust, with implementation of a preconditioning scheme and a detailed treatment of general fluid thermodynamics [31].

Owing to the demanding computational requirements of three-dimensional simulations, only a cylindrical sector with periodic boundary conditions in the azimuthal direction is simulated. The discrete orifices are converted into an axisymmetric slot through dynamic similarity. A multi-block domain decomposition technique, combined with a message passing interface for parallel computing, is applied to improve the computational efficiency. A typical simulation takes about 30,000 CPU hours on a single Intel Xeon processor to obtain statistically significant data, with a total time span of 30 ms, after reaching a fully developed state (~24 ms). The simulated data are subsampled every 30 time steps, with 1 μs time step intervals. According to the Nyquist criterion, a temporal resolution of 16.5 kHz is achieved.

### 1.4. High-Fidelity Simulation Results

Thirty high-fidelity simulation runs at design points defined by MaxPro were conducted. To isolate the effect of injector parameters, the mass flow rate for all runs is fixed at 0.15 kg/s. The first two design points designated by MaxPro are chosen as the baseline geometries, A and B in Table 2. The benchmark points used for assessing the accuracy of the emulator model are obtained by offsetting the design parameters of these two points. (Details are elaborated below, in the Results and Discussion section.)

Figures 3 and 4 show the instantaneous distributions of temperature and density for two neighboring design points, C and D in Table 2, selected to indicate different flow features in the design space. The key flow structures include a hollow gas core formed from the attachment of the swirling liquid film to the wall through centrifugal force, liquid accumulation near the injector headend due to the recirculation of LOX, and a conical liquid sheet spreading outward at the injector exit propelled by azimuthal momentum [25].

Table 2 Injector geometrics at design points colored blue in Figure 2.

|  | $L$ (mm) | $R_n$ (mm) | $\theta$ (°) | $\delta$ (mm) | $\Delta L$ (mm) |
|---|---|---|---|---|---|
| Design A | 20.0 | 3.22 | 52.9 | 0.52 | 3.42 |
| Design B | 41.9 | 3.05 | 65.5 | 1.57 | 1.00 |
| Design C | 43.1 | 5.00 | 70.0 | 0.50 | 2.79 |
| Design D | 37.7 | 2.82 | 45.8 | 1.17 | 3.80 |

Different flow physics, however, are also observed. The film thickness for design C is much thinner than for design D, with a larger spreading angle at the injector exit (34.6° compared to 29.2° for design D). Among the 30 design points, some act like swirling flows, as in design C, while others behave like jet flows (jet-like) as in design D. For convenience, the critical value of the spreading angle that separates swirling from jet-like flows is chosen to be 30°; this angle is considered to be the empirical indicator of whether the liquid stream has significant radial penetration in the downstream region. The 30 simulation runs are thus divided into two subgroups, swirl (spreading angle above 30°) and jet (spreading angle below 30°). In the next section, a machine-learning technique, decision tree, is introduced to identify the jet-swirl dichotomy. This is important because it directly influences the feature extraction and kriging processes described in the following sections.

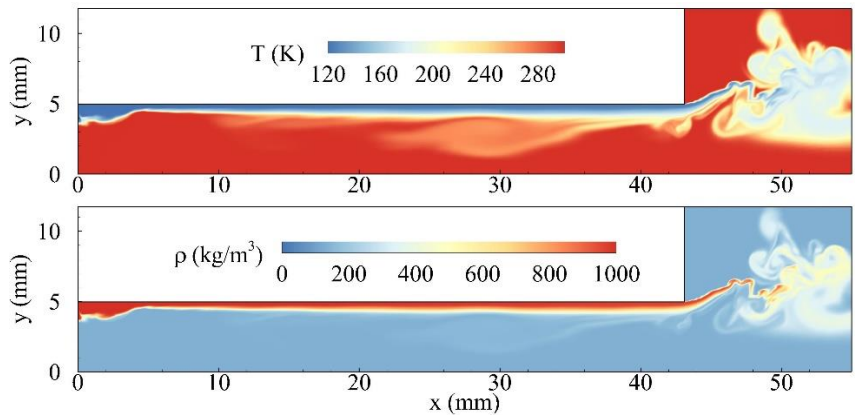

Figure 3. Instantaneous distributions of temperature and density for design C.

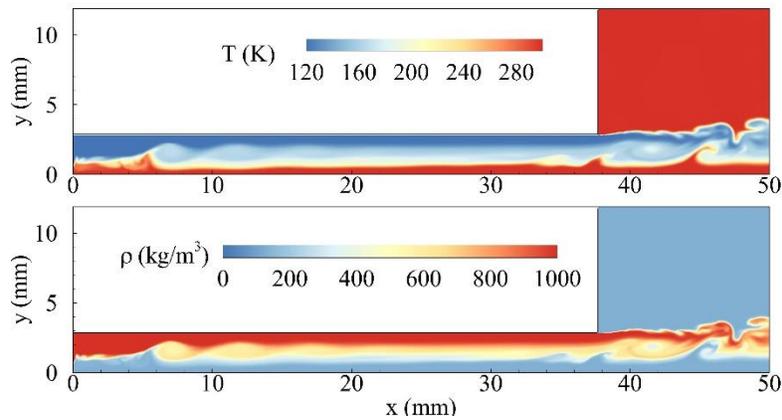

Figure 4. Instantaneous distributions of temperature and density for design D.

## 2. Data-Driven Framework

Design points may display similar or significantly different flow structures. In this section, the collected dataset from all 30 high-fidelity simulations is used to perform a data-driven analysis of the design space using machine-learning tools. The work consists of two components: (a) a sensitivity analysis for identifying important design parameters with respect to the quantities of interest, where the Sobol' indices [32] are herein, and (b) a decision-tree learning process with respect to the jet-swirl dichotomy, and the incorporation of this information into the emulator model. After the sensitivity analysis and decision tree learning, a technique called common POD (CPOD) is then implemented to extract the flow characteristics over the design space. Lastly, the time-coefficients for the obtained basis functions are used as training data for the kriging model. Integrating the results of this methodology, one can make accurate flow predictions at any new design setting. A flowchart of the overall data-driven emulator framework is provided in Figure 5.

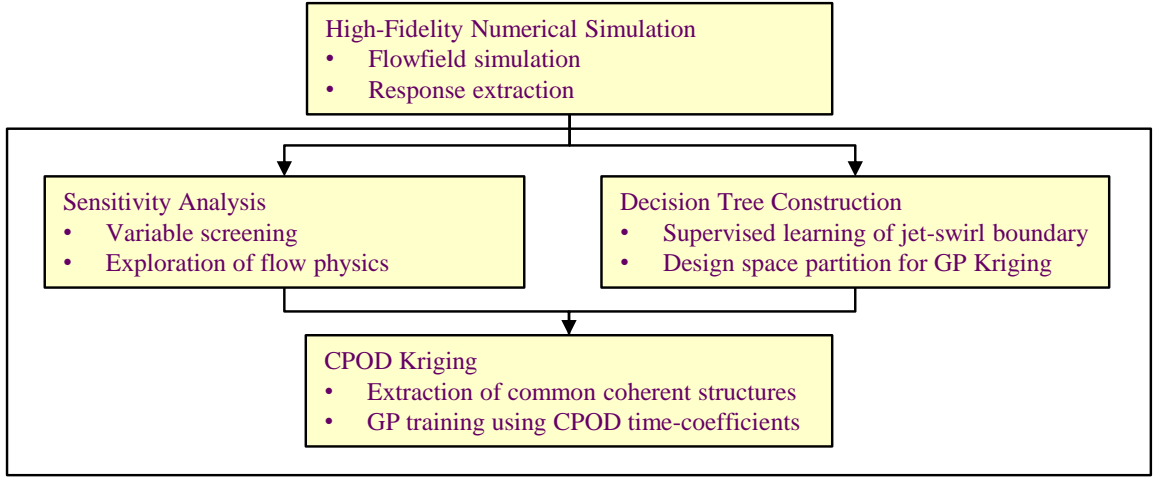

Figure 5. Flowchart for data-driven analysis and emulator construction.

## *2.1. Sensitivity Analysis*

The first component of this emulator framework is a sensitivity analysis using Sobol' indices [32] to identify which design parameters contribute more to changes in response outputs of interest, such as liquid-film thickness or spreading angle. The analysis is also a valuable tool for parameter reduction. The idea is to decompose the variations of certain desired output variables into the partial variation attributable to each input parameter and effects of interactions between parameters. This method of analyzing sensitivity has close connections to the classical analysis-of-variance employed in linear regression models [33].

To put it in mathematical terms, let $f(x)$ be the desired response output at design setting $x$, where $x = (x_1, x_2, \ldots, x_p)$ corresponds to the input parameters over a unit hypercube $[0,1]^p$. Specifically, for the current study, $p = 5$ and $x_1 = L, x_2 = R_n, x_3 =$ $\theta, x_4 = \delta$, and $x_5 = \Delta L$, with the design range for all parameters normalized to the interval $[0,1]$. Define the random variable $X$ as a uniform distribution over $[0,1]^p$, and let $f_0 = \mathbb{E}[f(X)]$ be the response mean and $D = \text{Var}[f(X)]$ be the response variance over the design range. The goal is to decompose the response variance $D$ into the contributions for each design parameter $x_1, \cdots, x_p$, as well as the effects of interactions between parameters. Consider the following decomposition:

$$f(x) = f_0 + \sum_{i=1}^{p} f_i(x_i) + \sum_{1 \leq i < j \leq p} f_{i,j}(x_i, x_j) + \cdots + f_{1,2,\cdots,p}(x_1, \cdots, x_p), \tag{1}$$

where each summand satisfies

$$\int_0^1 f_{i_1,\cdots,i_t}(x_{i_1}, \cdots, x_{i_t}) \, dx_k = 0, \tag{2}$$

for any $k = i_1, \ldots, i_t$ and has orthogonal components. In Eq. (1), the main effect index of input $i$ is

$$f_i(x_i) = \int (f(x) - f_0) \, dx_{-i}, \quad x_{-i} = \{x_1, \cdots, x_p\} \setminus \{x_i\}, \tag{3}$$

and the two-way interaction index of input $i$ and $j$ is:

$$f_{i,j}(x_i, x_j) = \int \{f(x) - f_0 - f_i(x_i) - f_j(x_j)\} \, dx_{-i,j},$$
$$x_{-i,j} = \{x_1, \cdots, x_p\} \setminus \{x_i, x_j\} \tag{4}$$

Squaring both sides of Eq. (1) and taking the integral over $[0,1]^p$, we get:

$$D = \sum_{i=1}^{p} D_i + \sum_{1 \leq i < j \leq p} D_{ij} + \sum_{1 \leq i < j < l \leq p} D_{ijl} + \cdots + D_{1,2,\cdots,p}, \tag{5}$$

where $D_u$ is the partial variance corresponding to a subset of parameters $u \subseteq \{1, \cdots, p\}$:

$$D_u = \int f_u^2(x_u) dx_u. \tag{6}$$

The Sobol' sensitivity indices for parameter subset $u$ can be defined as [32]:

$$S_u = \frac{D_u}{D} \in [0,1], \tag{7}$$

with larger values of $S_u$ indicating greater importance of the interaction effect for $u$.

In practice, Sobol' indices can be estimated as follows. First, a pseudo-random parameter sequence is generated using a low discrepancy Sobol' point set [34]. Second, this sequence is used to approximate the above integrals, which can then provide estimates for the corresponding Sobol' indices. The quantification of the response sensitivity for each parameter serves two purposes: (a) it provides a preliminary analysis of important effects in the system, which can guide further physical investigations, and (b) it allows for a reduction of the number of parameters that must be considered in the emulator, thereby providing a computationally efficient way to survey flow properties within the design space. A detailed discussion of this sensitivity analysis is presented in Section 3 for the current physical model.

### 2.2. Decision Tree

As mentioned in Section 1.4, there exists a jet-like/swirling flow dichotomy within the design space. For simulated design points, it is easy to classify whether such a parameter combination results in a jet-like or swirling flow, because the flowfield data are readily available. For design settings that have not been simulated, a data-driven technique is needed to make such a classification. There are two reasons why such a classification tool may be of interest. First, a boundary between jet-like and swirling cases can be established over the design space of interest, which can then be used to gain physical insight into the design space and to guide additional experiments. Second, the classification information can be used to train separate surrogate models within the jet-like and swirling domains. This partitioning of the emulator training dataset allows the model to extract different flow characteristics associated with jet-like and swirling behavior separately, and can thereby improve its predictive accuracy. A powerful machine-learning tool called 'decision tree' is employed for the classification process.

A decision tree is a decision support tool that models decisions and their possible consequences. Decision trees are one of the most popular predictive models in data mining and machine learning [35, 36]. Such methods are a part of a larger class of learning methods called supervised learning [37], which aims to predict an objective function from labeled training data. A classification tree, a special type of decision tree, is used here. It specializes in predicting classification outcomes, such as whether a parameter set has a jet-like or swirling flow. The trained model can be summarized by a binary tree, separating the design space into two subgroups. Each node of this tree represents a parameter decision, and each leaf of the tree indicates the class of outcomes, following the chain of decisions made from the tree root.

A classification tree can be trained using the following algorithm (see [37] for details). First, the simulated flowfields of each sampled design point are examined and classified as either jet-like or swirling flow, depending on the radial penetration of the propellant in the downstream region. Next, a search is conducted over all the design parameters and possible split-points, finding the parameter constraints which minimize misclassification. A branch is then made in the classification tree corresponding to the parameter constraint. The same branching procedure is repeated for each of the resulting child nodes. For the analysis in Section 4, the Gini impurity index [38] is selected as the misclassification measure:

$$\hat{p}_j(1 - \hat{p}_j) + \hat{p}_s(1 - \hat{p}_s), \tag{8}$$

where $\hat{p}_j$ and $\hat{p}_s$ are the proportions of jet-like and swirl cases in a split. The index measures how often a randomly chosen sample is incorrectly labeled when such a label is randomly assigned from the dataset. Notice that a Gini index of 0 indicates that (a) $\hat{p}_j = 1$ and $\hat{p}_s = 0$, or (b) $\hat{p}_j = 0$ and $\hat{p}_s = 1$, both of which suggest perfect classification.

This decision tree learning technique not only provides a means for partitioning the training dataset for the model into jet-like and swirling flows, but also reveals physical insights on the important design parameter constraints causing this jet-swirl dichotomy. The interpretability of these constraints is elaborated in Section 3.

## 2.3. Kriging Surrogate Model (Emulator)

The primary objective of this work is to develop an emulator model that uses the existing information of 30 simulation runs to predict the flowfield of a new design point within a practical turnaround time. With these tools in hand – the sensitivity analysis for parameter screening and the decision tree for partitioning the design space into jet-swirl cases – a surrogate model for flowfield emulation is proposed. The kriging surrogate model, also known as an emulator, combines machine-learning techniques, statistical modeling, and a physics-driven data reduction method. A brief motivation for each part of this model is provided, before delving into the specific mathematical details.

First, the proposed model is constructed through a POD analysis of the simulation dataset used for training. For a given flow property, the POD analysis determines a set of orthogonal basis functions, $\phi_j$, such that the projection of the property onto these basis functions has the smallest error, defined as $E\left(\|f - \hat{f}\|^2\right)$ where $E(\cdot)$ and $\|\cdot\|$ denote the time average and norm in the $L^2$ space, respectively [20]:

$$\hat{f}(\pmb{x}, t) = \sum_{j=1}^{n} \beta_j(t) \phi_j(\pmb{x}) \tag{9}$$

The basis functions, or mode shapes, are spatial distributions of the fluctuating fields of the flow properties, which can be closely linked to physical phenomena and coherent structures. The basis functions are ordered in such a way that the lowest modes have the highest "energy," as defined by the inner product of $f$. The flow properties for POD analysis include pressure, density, temperature, and velocity components. POD decomposition yields not only the eigenfunction modes, $\phi_j$, and but also their corresponding time-varying coefficients, $\beta_j$, which are referred to as POD coefficients.

While the usage of POD simplifies the complex nature of a spatio-temporal model, a common set of basis functions is required for the emulator. Physically, this means a common set of coherent structures needs to be extracted over the design space. One option is to select a computational region of interest that is unaffected by any design changes [39]. Taking advantage of the basis functions generated by the POD analysis, an emulator can be obtained as long as a set of common basis functions exist. One of the challenges for the current study is the wide disparity of geometries in the design space, as shown in Figure 6. The present study utilizes a common grid for the 30 grid systems to find a set of common basis functions. To achieve this, the densest grid system among all cases is identified and split into four sections covering the effects of design parameters on the simulated grid. This partitioned grid is used for interpolation and rescaling of each simulated case to obtain a common grid. Then, an inverse distance weighting interpolation method with ten-nearest neighborhood points is used to map the original raw data onto the common grid [40]. Algorithmically, the CPOD expansion is obtained by first rescaling the different cases to the common grid, then computing the POD expansion, and finally rescaling the resulting modes back to the original grid.

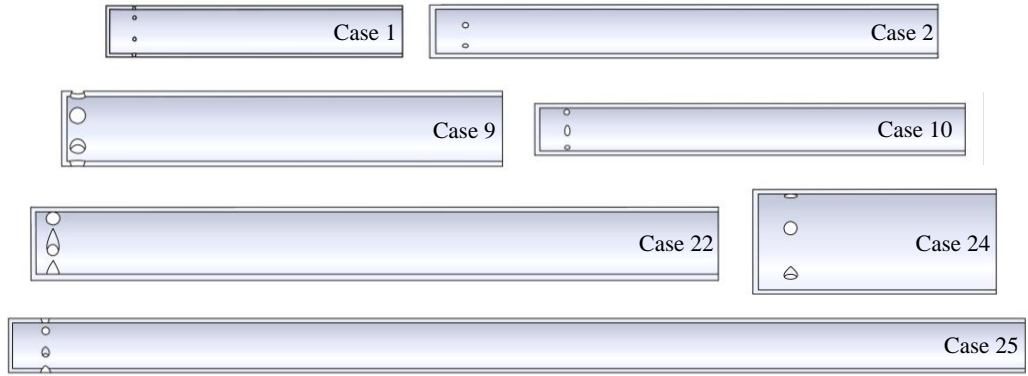

Figure 6. Schematics of different injector geometries in the design space.

Because of the limited variation of the Reynolds number among the different injector geometries, the scaling of the data to the common grid is appropriate in the present study of nonreacting flows. The smallest injector diameter of concern is 4 mm, with a corresponding exit velocity of 27.5 m/s. With the liquid oxygen density being 1000 $kg/m^3$ and viscosity 0.114 cP, the Reynolds number based on the injector diameter is about $9.6 \times 10^5$. The largest injector diameter in the design space has a value of 10 mm, and the corresponding exit velocity is 11 m/s. At the same operating condition, the Reynolds number comes out to about $9 \times 10^5$. For some geometries where the liquid film does not produce a noticeable spreading angle, the Reynolds number is reduced to about $9 \times 10^5$. Despite this difference, the model is capable of avoiding excessive smoothing, provided the correlation function is bounded correctly. Such scaling of POD modes to establish common basis functions is vital to the emulator. It should be noted that this scaling is only appropriate for flow simulations that do not exhibit distinctively different physical phenomena. Additional similarity parameters may be necessary when different physics and chemical reactions are incorporated, as reviewed by Dexter et al. [41].

The mathematical details for CPOD are provided below. Suppose $n$ simulations are conducted at various design geometries $c_1, \ldots, c_n$ and let $f(x, t; c_i)$ be the simulated flowfield at design $c_i$ for a given time $t$ and spatial coordinate $x$. The $k$-th CPOD mode is defined as

$$\phi_k(x) = \underset{\psi:\|\psi\|_2=1}{\mathrm{argmax}} \sum_{i=1}^{n} \int \left[ \int \mathcal{M}_i[\psi(x)] f(x, t; c_i) dx \right]^2 dt, \quad (10)$$

$$s.t. \int \psi(x) \phi_l(x) dx = 0, \forall l < k$$

Here, the map $\mathcal{M}_i: \mathbb{R}^2 \to \mathbb{R}^2$ is the transformation which linearly scales spatial features from the common geometry $c$ to the $i$-th geometry $c_i$. The sequence of POD coefficients is defined as:

$$\beta_k(c_i, t) = \int \mathcal{M}_i\{\phi_k(x)\} f(x, t; c_i) dx, \quad (11)$$

with the corresponding POD expansion using $K$ modes given by:

$$f^{(K)}(x, t; c_i) = \sum_{k=1}^{K} \beta_k(c_i, t) \mathcal{M}_i\{\phi_k(x)\}. \quad (12)$$

The transformation allows for the extraction of common basis functions. In addition, the obtained modes can be used to identify key mechanisms of flow dynamics. It should be noted that reacting-flow simulations are characterized by additional dimensionless parameters and linear mapping may not perform well when combustion is involved.

Two computational challenges need to be addressed to implement this methodology. As previously mentioned, to calculate the inner product of the snapshots from different simulation cases, a common set of spatial grid points is needed. Not only does the inner product calculation become the computation bottle-neck, the number of modes required to capture a certain energy level is significantly increased relative to an individual simulation. The computation of CPOD modes and associated time-varying coefficients requires eigen-decomposition of a $nT \times nT$ matrix, where $n$ is the number of simulation cases and $T$ is the number of snapshots. This usually requires $O(n^3 T^3)$ computation work. An iterative method of eigen-decomposition based on periodic restarts of Arnoldi decompositions is then used to quickly calculate the first few eigenvectors with the largest eigenvalues. These eigenvalues can also be interpreted as the amount of the "energy" as defined by the inner product used to calculate the covariance matrix. For a particular reconstruction using a linear combination of POD modes and associated time-varying coefficients, there is reconstruction error, which decreases when more eigenvectors, the POD modes, are included.

Next, a kriging model is applied to the CPOD time-varying coefficients $\beta_k(\boldsymbol{c}_i, t)$. Kriging is a powerful machine-learning tool which, using function values sampled at a set of input parameters, can approximate well the entire function surface over its domain. The driving force behind kriging is the usage of a GP governed by a preset covariance function. With appropriately chosen parameters, the kriging model provides the best linear unbiased estimator of the responses at designs that have not been simulated [10]. Furthermore, the resulting posterior distribution of this prediction will also be Gaussian. With the mean and variance computable in closed form, uncertainty quantification and confidence intervals can be calculated easily. Kriging (and more generally, GP-based learning) has been applied to great success in a variety of fields [42].

The mathematical approach of kriging is described here. For notational simplicity, let $\beta(\boldsymbol{c})$ denote $\beta_k(\boldsymbol{c},t)$, the $k$-th CPOD coefficient at setting $\boldsymbol{c}$ and time-step $t$. As the temporal resolution is fine, there is no practical need to estimate temporal correlations, especially because predictions will not be made in between time-steps. This time-independent emulator uses independent kriging models at each instant of time, assuming the following GP model:

$$\beta(\boldsymbol{c}) = \mu + Z(\boldsymbol{c}), \qquad Z(\boldsymbol{c}) \sim N\{0, \sigma^2 R(\cdot,\cdot)\} \tag{13}$$

Here, $\mu$ is the mean, $Z(\boldsymbol{c})$ is a zero-mean GP with variance $\sigma^2$, and $R(\cdot,\cdot)$ is a pre-specified correlation function governed by unknown parameters $\eta's$. A typical choice for $R(\cdot,\cdot)$ is the so-called Gaussian correlation function:

$$R(\boldsymbol{c}_i, \boldsymbol{c}_j) = \exp\left[-\sum_{k=1}^{p} \eta_k (c_{ik} - c_{jk})^2\right] \tag{14}$$

where $p$ is the number of input parameters.

Now, suppose the function values $\boldsymbol{\beta}^{(n)} = [\beta(\boldsymbol{c}_i)]_{i=1}^n$ are observed at input settings $\{\boldsymbol{c}_i\}_{i=1}^n$, and let $\boldsymbol{c}_{new}$ be a new setting for which prediction is desired. Conditional on the observed values $\boldsymbol{\beta}^{(n)}$, the best linear unbiased estimator of $\beta(\boldsymbol{c}_{new})$ can be shown to be [10]:

$$\hat{\beta}(\boldsymbol{c}_{new}) = \mu + \boldsymbol{r}_{new}^T \boldsymbol{R}^{-1}(\boldsymbol{\beta}^{(n)} - \mu \boldsymbol{1}) \qquad (15)$$

Here, $\boldsymbol{1}$ is the $n$ x 1 vector of ones, $\boldsymbol{r}_{new} = [R(\boldsymbol{c}_i, \boldsymbol{c}_{new})]_{i=1}^n$ is the $n$ x 1 vector of correlations between the new point and sampled points, and $\boldsymbol{R} = [R(\boldsymbol{c}_i, \boldsymbol{c}_j)]_{i=1, j=1}^{n \quad n}$ is the covariance matrix for the sampled points. It can be shown that such a predictor minimizes the mean-squared prediction error (MSPE), a commonly-used criterion for prediction error. In the context of flowfield prediction, employing this kriging estimator allows us to obtain accurate flow predictions from the CPOD coefficients. It can also be shown [10] that this best MSPE predictor is unbiased, matching the expected and true function value.

To close the formulations, the model parameters $\mu$, $\sigma^2$ and $\eta$ need to be trained using data. A technique called maximum likelihood estimation (MLE), a ubiquitous estimation technique in statistical literature [43], is employed. The key idea in MLE is to search for the optimal parameter setting that minimizes the likelihood function of the GP model. In the present work, optimization is achieved by means of the L-BFGS algorithm [44], a method employed for many training algorithms. A more detailed explanation can be found in Santner et al. [10].

The kriging models are trained independently over each time step, due to the inherent fine-scale temporal resolution of the simulation. This time-independence assumption is made for two reasons. First, the fully developed flow is treated as statistically stationary and has high frequency resolution, so there is no practical value for estimating temporal correlations. Second, as in the high-fidelity simulation procedure, the assumption of time-independence allows exploitation of parallel computation in training the emulator model. Once the model is trained, the predictor is used with the CPOD expansion to predict the flow evolution at a new design point, i.e.,

$$\hat{f}(\boldsymbol{x}, t; \boldsymbol{c}_{new}) = \sum_{k=1}^{K} \hat{\beta}_k(\boldsymbol{c}_{new}, t) \mathcal{M}_i\{\phi_k(\boldsymbol{x})\} \tag{16}$$

It is worth noting that the computation time of the proposed model is orders of magnitude smaller than that of LES. Simulation data that typically takes a week, or around 30,000 CPU hours to simulate, can be predicted by the model with an associated uncertainty in a manner of tens of minutes. The full emulator model and algorithm are provided in the complementary statistical paper [23], which considers the statistical properties of a broader class of models. The current paper focuses on applying new machine-learning techniques and investigates the practical performance of the emulator with respect to flow physics.

## 3. Results and Discussion

### *3.1. Sensitivity of injector geometrical parameters*

Liquid-film thickness and spreading angle are two important injector characteristics. Inviscid, incompressible flow theory predicts the spreading angle as a function solely of the geometric constant [24, 45], and it increases with increasing geometric constant. For real fluids at supercritical conditions as treated in the present study, the density gradient represents the transition region [25, 26]. The spreading angle can be determined based on the slope of the maximum density gradient near the injector exit. The film thickness is measured by its radial position.

To gauge the importance of each injector parameter on the liquid-film thickness and spreading angle, a sensitivity analysis using a Monte Carlo estimate of Sobol' indices was performed [32]. Figure 7 shows the primary effects from this sensitivity analysis. The points indicate the Sobol' index estimate for each design parameter, with lines indicating the Monte Carlo integration error for each index estimate. The lines were calculated based on a 95% confidence interval of the estimate. The slot width ($\delta$) is found to be the parameter with the largest Sobol' index and thus the strongest influence on the spreading angle. Physically, this can be explained by how geometric parameters govern the inlet flow properties. Assuming a constant mass flow rate, the incoming velocity is directly proportional to the slot width, and a decrease in slot width increases liquid-film momentum, which gives the liquid film increased momentum.

Similarly, the tangential inlet angle ($\theta$) and the slot width significantly affect the liquid-film thickness, while the length ($L$) and radius ($R_n$) of the injector have minor effects. The tangential inlet angle controls the direction of momentum. As the angle increases, more azimuthal momentum is imparted to the liquid film, increasing the spreading angle as the propellant expands radially downstream. The length and radius can dictate how much viscous loss is experienced by the propellant, as it travels in both the axial and azimuthal directions. The present study, however, has shown viscous losses to be a minor effect.

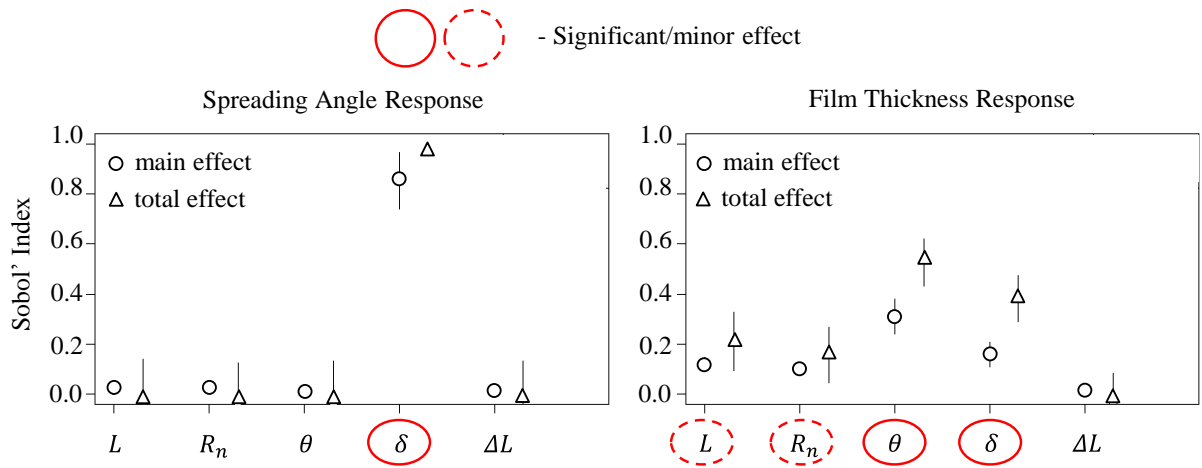

Figure 7. Sensitivity analysis of liquid-film thickness and spreading angle.

Figure 8 shows the two-factor interaction effects. It further demonstrates that the main design parameters are the slot width and the tangential inlet angle, which couple to affect the liquid-film response. This is hardly surprising, as slot width and inlet angle govern flow area and direction of momentum, respectively. The mass and momentum conservation equations are inherently coupled to govern the flowfield.

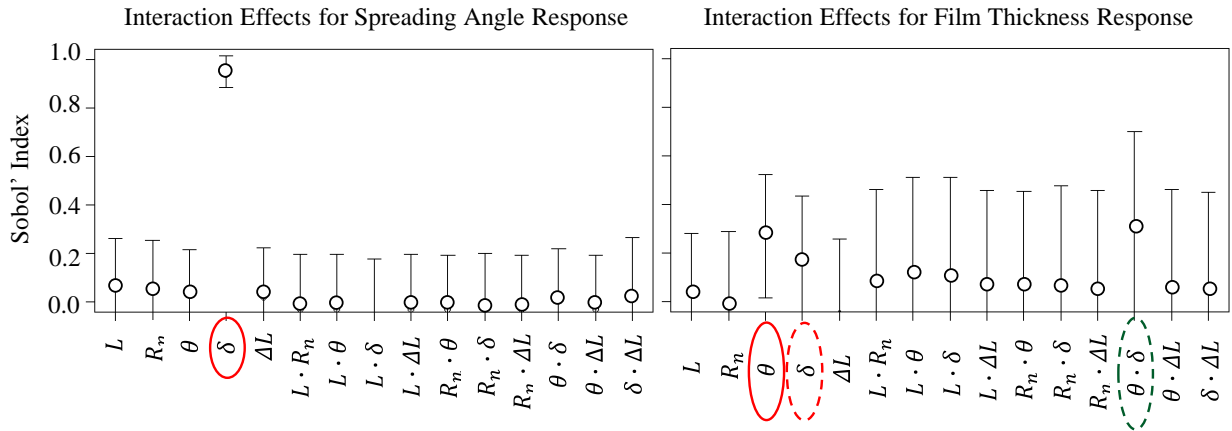

Figure 8. Two-factor interaction of liquid-film thickness and spreading angle.

As previously mentioned, the empirical geometric constant for a swirl injector can be employed to estimate the film thickness and spreading angle, using the hydrodynamics theories described by Bazarov and Yang [24, 45]. These theories, however, are based on the assumption of incompressible, inviscid flows and can only be used as a preliminary guide. In real injectors, viscous and compressibility effects must be considered. The liquid viscosity results in boundary layer formation along the walls, causing spatially non-uniform velocity profiles. A primary effect of compressibility lies in the existence of acoustic waves [26, 46]. The supercritical conditions within high pressure systems make these effects even more pronounced. High-fidelity simulations taking into account real-fluid effects are required to address these issues [26, 31, 46].

### *3.2. Decision tree exploration of injector design space*

With further examination of the simulated design points, there is a clear distinction between two types of underlying physics. One type is the expected swirling film that noticeably spreads radially upon exiting the injector. The other type is a jet-like behavior of the liquid film where the radial spreading is weak. The DoE methodology utilizes space-filling properties such that design points in both regimes are simulated. This section explores how to efficiently incorporate this information into the CPOD methodology to refine prediction results.

Design A and B (geometric parameters are listed in Table 2) are each chosen as baseline geometry for determining off-design points. By offsetting injector parameters, two benchmark design points are obtained (denoted as red points in Figure 2). Design A is classified with swirling behavior. Although design B is classified with jet-like behavior in its developing stage, the flowfield transitions to a swirling flow in its stationary state. This trend may be an indicator of Design B being near the jet-swirl regime boundary.

A full design trade-off study requires quantifying how every parameter affects key performance metrics. Hence, all injector variables are retained for the first benchmark, Benchmark E. The second benchmark, Benchmark F, only varies design parameters with significant effects on the liquid-film response. The corresponding geometries are shown in Table 3. For Benchmark E, each design parameter has a +10% deviation away from those of Design A. With normalized parameters, the distance traversed in the design space is estimated to be about 18.1%, as calculated in the $L_2$ linear sense.

Table 3. Injector geometries for benchmark cases.

|  | $L$ | $R_n$ | $\theta$ | $\delta$ | $\Delta L$ |
|---|---|---|---|---|---|
| Benchmark E | 22.0 mm | 3.22 mm | 58.2° | 0.576 mm | 3.42 mm |
| Benchmark F | 37.7 mm | 3.06 mm | 59.0° | 1.417 mm | 1.00 mm |

The sensitivity study showed that the injector radius and the injection location have less of an effect than the slot width and tangential inlet angle on the film thickness and spreading angle. They are thus fixed, and the other three parameters are offset from Design B by -10% to explore the design space at Benchmark F. The closest two simulation points are Designs C and D. The neighboring points are provided because Design B seems to be near the jet-swirl dichotomy.

The second component of the data-driven framework for the design survey is a decision tree [37]. Figure 9 shows the decision-tree splitting process, indicating how the algorithm decides the way an injector parameter dictates whether the flow is jet-like or swirling. The initial decision between the two behaviors is achieved by assessing the extent to which the liquid film spreads radially from the injector exit. The numeric outputs are essentially binary flags between the two subgroup classifications. Intuitively, when the tangential inlet angle, $\theta$, is smaller, there is less azimuthal momentum in the liquid film to cause radial spreading. When the injector inlet, $\delta$, becomes large, the decreased momentum results in jet-like behavior. The decision tree quantifies these effects and predicts a jet-like injector with $\theta < 60.02°$ and $\delta > 1.40$ mm. Following the previous two criteria, if the tangential inlet angle is large enough, that is, $\theta > 49.24°$, the injector retains swirling behavior.

The two benchmark cases are used to verify the decision tree. With such an algorithm, simulation results can be predicted using the model with proper training data. As the next section will further detail, the emulator relies upon the set of common basis functions extracted from the dataset. With two types of underlying coherent structures, the two datasets should be trained separately to predict design parameter sets that lead to their corresponding flow behavior.

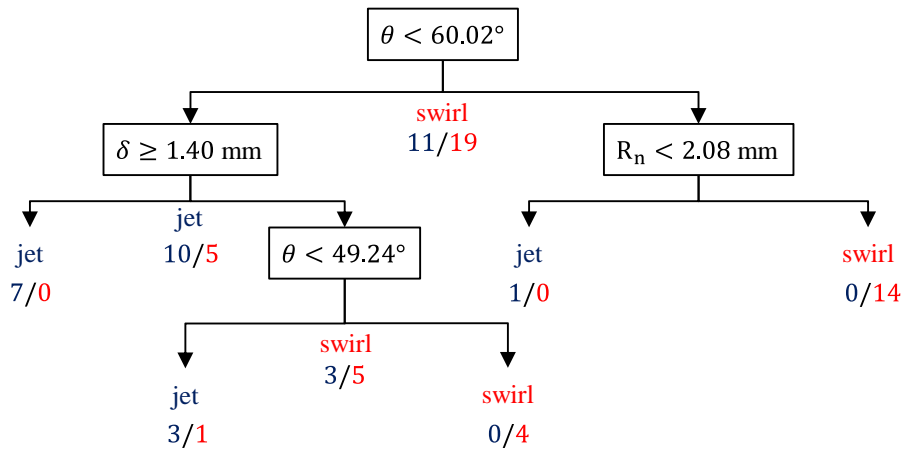

Figure 9. Decision tree splitting process with numeric classifiers.

### *3.3. Surrogate Model*

In order to train an emulator and make predictions, a set of common basis functions must be utilized as previously mentioned. Figure 10 shows the process of the common grid generation. The red lines partition the axisymmetric domain for each case into five regions: injector headend region, injector interior, and three subregions downstream of the injector. The densest grid system among the 30 training cases is selected as the common grid, upon which the partitioned regions for all other cases are then scaled to the corresponding regions in the common grid. Moreover, the original data is interpolated with an inverse distance weighting interpolation method using the ten nearest neighborhood points. The results on the common grid are used for POD analysis.

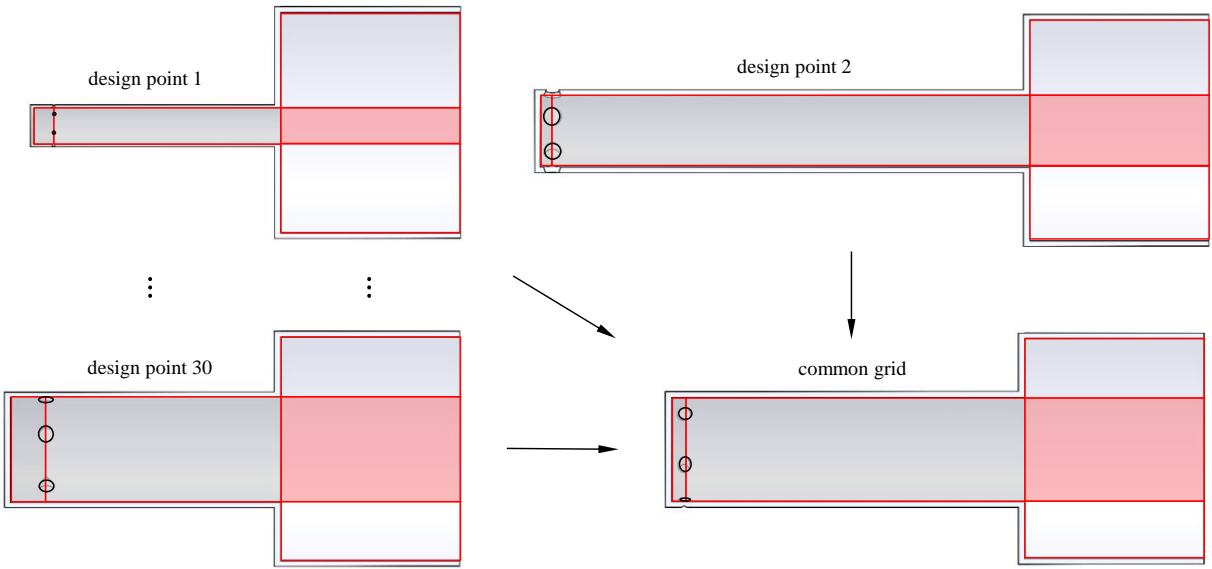

Figure 10. Schematic of common grid generation process.

Figure 11 shows the energy spectrum of the azimuthal velocity captured by the CPOD analysis. 45 CPOD modes are required to retain 99% of the energy and limit the corresponding truncation error for the reconstruction. The leading two modes are presented in Figure 12, both indicating swirling flow structures with dominant fluctuations near the injector wall. The flow evolution within the injector and subsequent liquid-film development downstream of the exit are clearly observed.

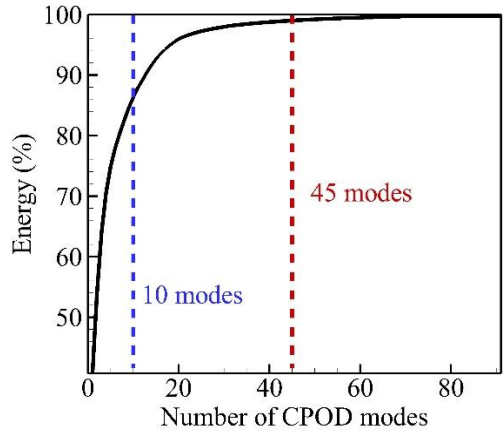

Figure 11. Energy spectrum of CPOD modes for azimuthal velocity component.

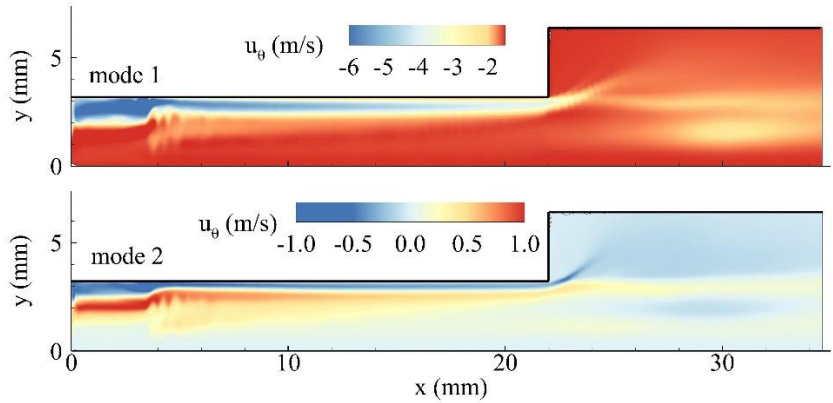

Figure 12. First two CPOD modes of azimuthal velocity.

The kriging of time-varying coefficients combined with the CPOD modes allows for emulation of the spatio-temporal evolving flow at a new design point. The CPOD modes represent the common physics extracted from the training dataset and we assume that the new injector geometry contains similar physics, and thus a similar POD deconstruction. Figure 13 shows a comparison of the simulation and emulation results, an instantaneous temperature distribution snapshot, of Benchmark E ($L = 22.0\ mm$, $R_n = 3.22\ mm$, $\theta = 58.2°$, $\delta = 0.576\ mm$, and $\Delta L = 3.42\ mm$). Good agreement is obtained, illustrating the same qualitative trends for the flow structures, with liquid film along the injector wall and a center recirculating flow downstream of injector. The POD analysis can be interpreted as a spatial averaging technique using the covariance matrix of the flow variable of interest. Some flow details, such as the surface wave propagation of the liquid film, may be smoothed out due to averaging. This concern, however, can be addressed effectively using the aforementioned statistical and optimization algorithms to tune GP model parameters. The resultant emulator model thus mitigates the smoothing effects and captures the flow structures well.

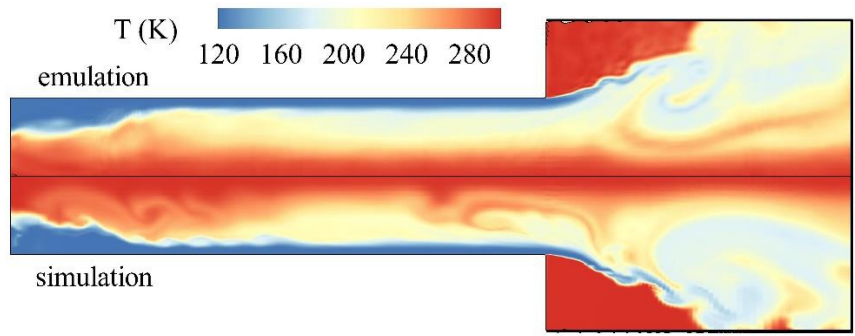

Figure 13. Comparison of instantaneous temperature distributions for Benchmark E.

### 3.4. Response Performance Metrics

As a preliminary comparison, a kriging surrogate model was applied to the extracted liquid-film thickness and spreading angle at the injector exit. The training process was implemented for the 30-case dataset. The following discussion is based on benchmark E: a swirl case. The liquid-film thickness is estimated, based on hydrodynamics theories, to be 0.618 mm, and the spreading angle 91.8°. The point-wise scalar emulator predicts a liquid-film thickness of 0.520 mm and a spreading angle of 99.0°. The data are compared with the simulation results of 0.430 mm and 103°, respectively. Figure 14 shows the variation of the film thickness along the injector wall.

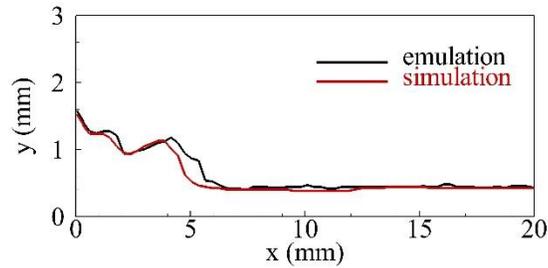

Figure 14. Comparison of mean liquid-film thickness along axial distance.

The average error is 6.74%. At the injector exit, the extracted scalar prediction results achieve a 20.9% error for the liquid-film thickness and a 3.88% error for the spreading angle. The time-average predicted film thickness and spreading angle obtained were 0.420 mm and 107°, corresponding to percentage errors of 2.38% and 3.88% respectively. The spreading angle prediction improved by a percentage for this geometry. The model matches the simulation in terms of key features such as the liquid-film location/thickness and spreading angle, performance measures needed for an injector design trade-off study. In addition, this reduced basis model is able to capture the flow evolution.

For Benchmark F, the baseline case (Design B) develops from a jet-like behavior into a swirling behavior, as shown in Figure 15. The implication is that its design parameters are near a critical hyperplane separating jet-like and swirling flow.

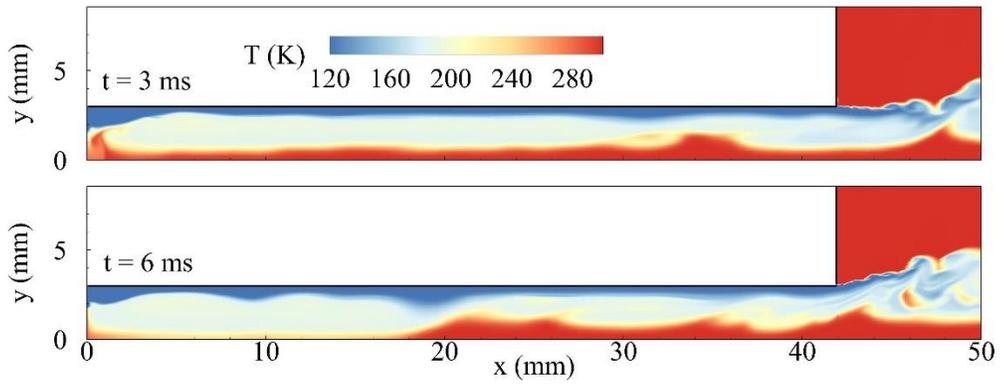

Figure 15. Baseline case for benchmark F.

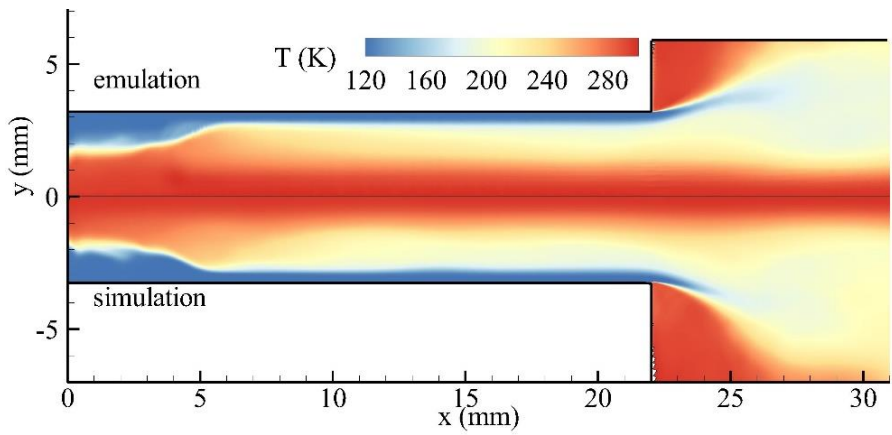

a) Swirl-like case

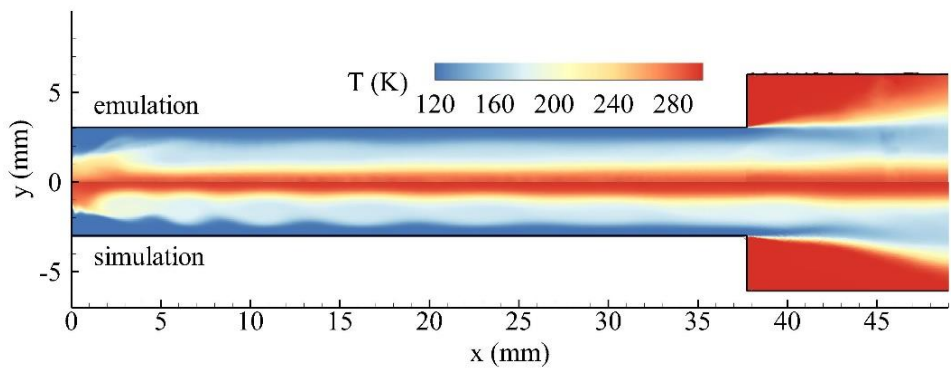

b) Jet-like case

Figure 16. Mean temperature distributions for benchmark cases a) swirl-like case and b) jet-like case.

Figure 16 shows the mean temperature distributions for the two benchmark cases. The figure a) shows emulation (top) and simulation (bottom) for benchmark E, which produces a swirling flow. The accumulation of liquid propellant at the injector headend is observed in both results, although the prediction worsens in the upstream section relative to the downstream. The downstream liquid-film thickness and spreading angle match well.  Figure b) shows Benchmark F, which produces a jet-like flow. A standing wave appears in the upstream portion of the injector for this case. The emulation result captures the wavy structure only to some extent, but the downstream flow structures, for liquid-film thickness and spreading angle, are better predicted. In the region where the film breaks apart, less propellant appears in the simulation result.

### 3.5. Root Mean Square Relative Error

The root mean square relative error (RMSRE) as defined by

$$RMSRE(t; S) = \frac{[\int_S \{f(\mathbf{x}, t; \mathbf{c}_{new}) - \hat{f}(\mathbf{x}, t; \mathbf{c}_{new})\}^2 d\mathbf{x}/|S|\,]^{1/2}}{\max(f(\mathbf{x}, t; \mathbf{c}_{new})) - \min(f(\mathbf{x}, t; \mathbf{c}_{new}))} \times 100\%, \qquad (17)$$

where $S$ is the desired region, $|S|$ is the number of gridpoints under $S$, $f(\mathbf{x}, t; \mathbf{c}_{new})$ is the simulated flowfield at geometry $\mathbf{c}_{new}$, $\hat{f}(\mathbf{x}, t; \mathbf{c}_{new})$ is the emulated flowfield, and $\max(f(\mathbf{x}, t; \mathbf{c}_{new}))$ and $\min(f(\mathbf{x}, t; \mathbf{c}_{new}))$ are the maximum and minimum values of $f(\mathbf{x}, t; \mathbf{c}_{new})$ over $\mathbf{x}$, respectively.

The numeric comparisons are tabulated in Table 4 as the RMSRE results. There are minor discrepancies near the injector wall, and the error is reduced with prediction results cut off at the injector exit.

Table 4. RMSRE temperature distribution results.

| Benchmark Case | Overall | Upstream | Downstream |
|---|---|---|---|
| E (swirl) | 5.18% | 6.62% | 3.10% |
| F (jet) | 8.65% | 8.30% | 9.03% |

Figure 17 shows the axial variation of the temperature distribution in the radial direction for Benchmark E. The mean flow development in the axial direction is captured as the transition region is matched. There is a deviation near the injector wall and the sharp temperature gradient experienced by the simulation result is smoothed by the emulator. Similar results are obtained for the Benchmark F, which are not shown here.

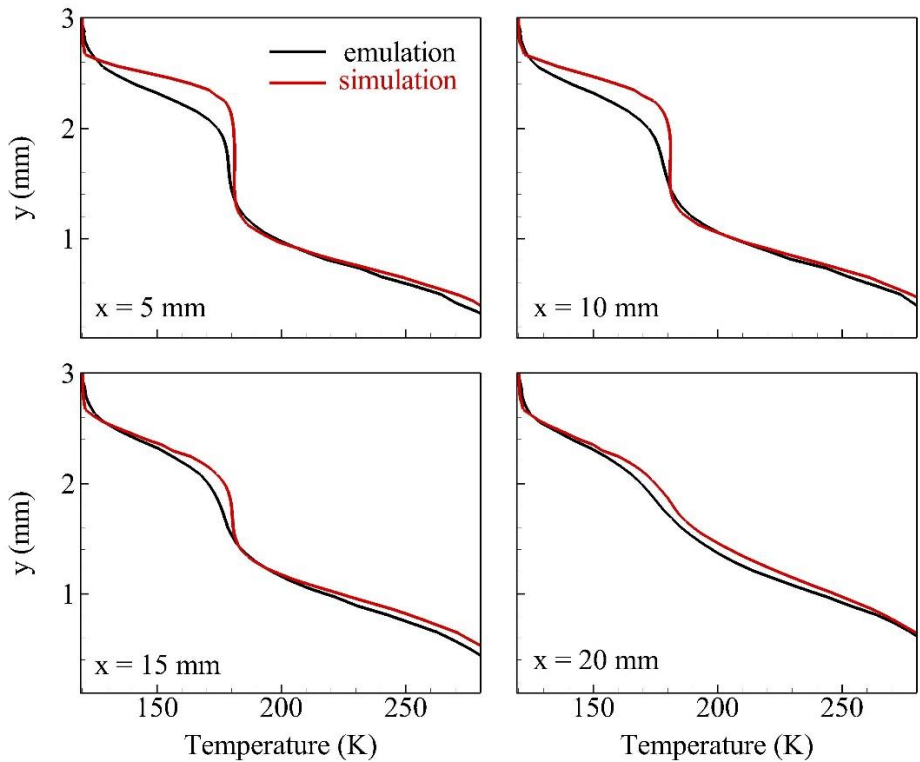

Figure 17. Axial variation of temperature distribution in radial direction.

To reiterate the importance of the decision tree, a comparison is made with predictions from the emulator without dataset classification [23]. Although Benchmark E experiences marginal improvement, Benchmark F's prediction is visibly enhanced.

Table 5. RMSRE temperature distribution results (without dataset classification).

| Benchmark Case | Overall | Upstream | Downstream |
|---|---|---|---|
| E (swirl) | 5.93% | 6.70% | 5.09% |
| F (jet) | 13.2% | 7.43% | 17.7% |

To demonstrate that the emulator is also capable of modeling other flowfield variables, predictions of the axial velocity are made. Figure 18 shows a contour plot comparison for Benchmark E. The recirculating gas core is observed, and the convection speeds within the injector match. Both the gaseous core and the swirling film are predicted well, as axial velocity RMSRE results in Table 6 outperform temperature results.

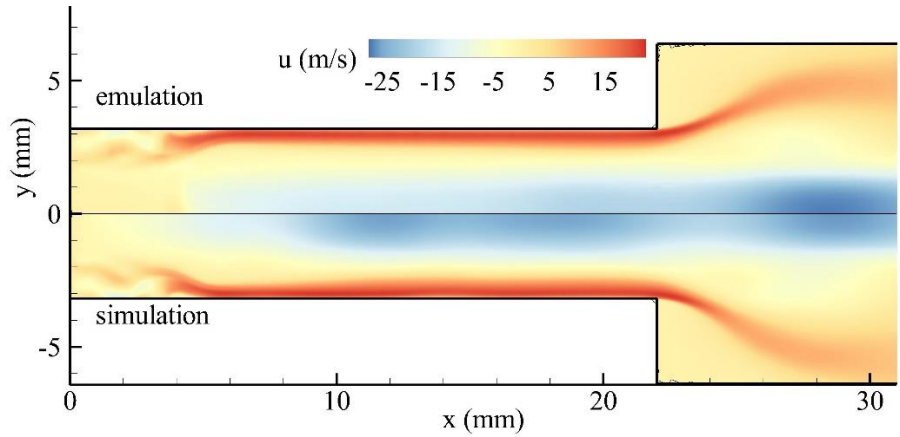

Figure 18. Mean axial velocity distribution for benchmark E.

Table 6. RMSRE velocity distribution results.

| Benchmark Case | Overall | Upstream | Downstream |
| --- | --- | --- | --- |
| E (swirl) | 4.12% | 4.58% | 3.64% |
| F (jet) | 3.97% | 4.71% | 2.85% |

Figure 19 shows the variation of the axial velocity distribution in the radial direction through the injector for Benchmark E. The mean flow development in the axial direction is captured as the transition region is matched. There is a deviation near the injector centerline and the sharp axial gradient experienced by the simulation result is smoothed by the emulator. Similar results are obtained for Benchmark F, which are not shown here.

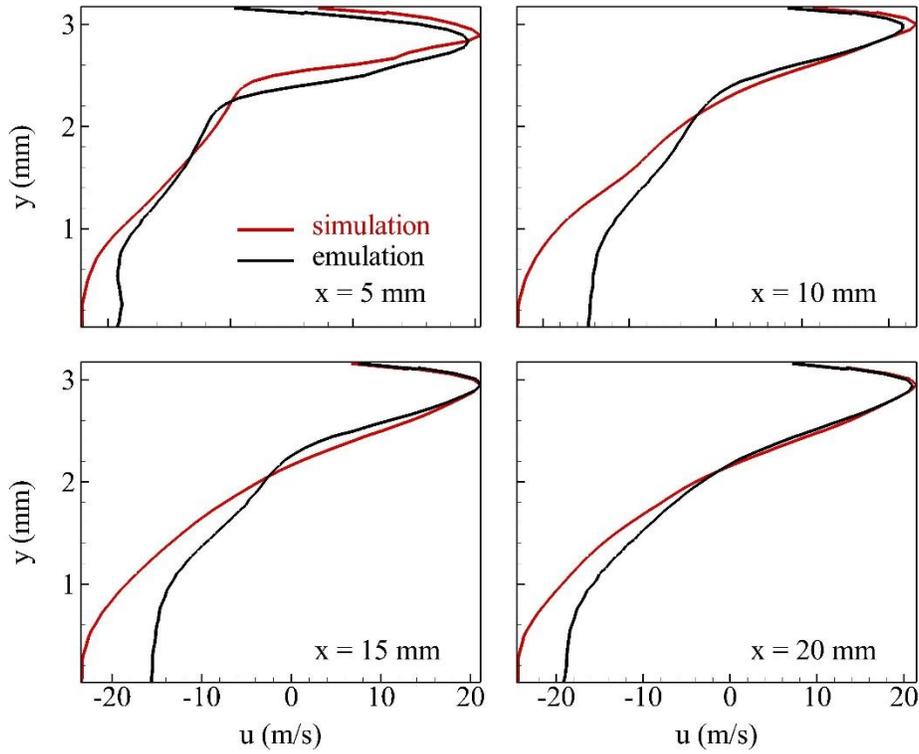

Figure 19. Axial variation of velocity distribution in the radial direction.

## 3.6. Injector Dynamics

The injector dynamics involved in this example problem cover a wide range of time and length scales. The rich set of physics can be quantified using a power spectral density (PSD) analysis. Figure 20 shows the position of the pressure probes within the fluid transition region used for the analysis. The probes are located near the film surface to capture the flow oscillations as the swirling fluid exits the injector.

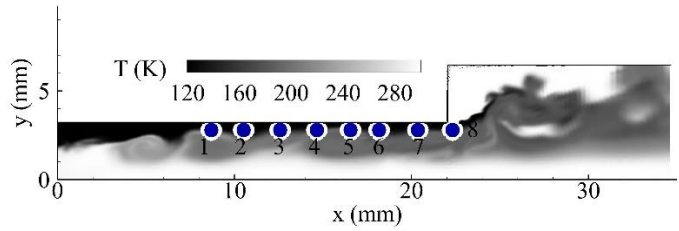

Figure 20. Probe positions along liquid film surface.

Injector dynamics involves downstream pressure fluctuations causing pressure drop oscillations across the liquid film, which in turn triggers mass flow rate variations. A spectral analysis can quantify these oscillations and capture the periodicity of the flow structures. Mathematically, the PSD can be interpreted as the Fourier transform of the autocorrelation function for a signal. The pressure PSDs were calculated for both the simulation and surrogate model results. Figure 21 shows the PSD results of probes 1, 3, 5 and 7; the frequency content is observed to be well quantified.

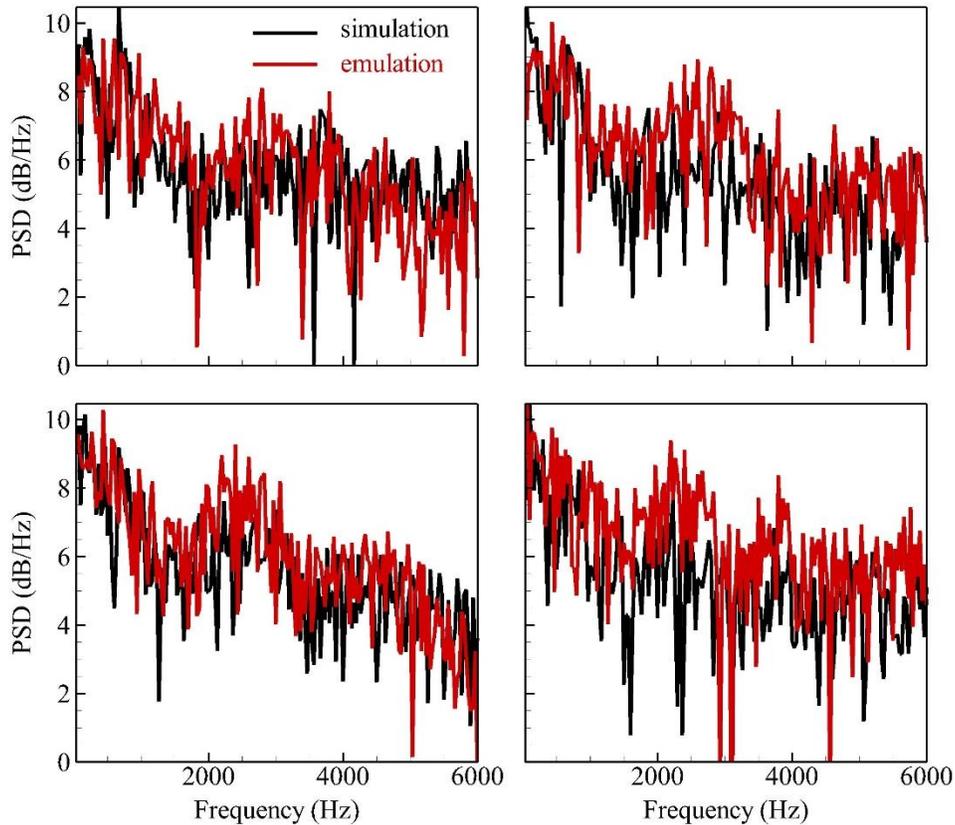

Figure 21. PSD results of pressure fluctuations for probes 1, 3, 5, and 7.

The high-frequency modes with small wavelengths that are typically present in swirl injectors having a vortex chamber are not prominent. Most of the signal is comprised of low-frequency modes with long wavelengths that transmit through the liquid film and exit the injector. This is usually in the form of surface wave propagation of the film, a longitudinal hydrodynamic instability. Disturbances in the chamber can propagate back into the injector through this mechanism and induce pressure drop changes across the tangential inlets, which in turn produce a mass flow rate fluctuation [24, 45]. Downstream of the injector exit, the dynamics become more broadband and no dominant oscillations can be found, due to the strong interactions between the shear layer and the recirculation zone generated from vortex breakdown. In addition, acoustic waves propagate, couple and interact with hydrodynamic waves, appearing as several different frequencies within the spectral content. The probes show similar dynamics in both the simulation and emulation results. This analysis is an indicator that the emulator is capable of modeling the periodic oscillations of the simulation results, properly demonstrating the flow dynamics.

*3.7. Uncertainty Quantification*

The emulator model also allows for the quantification of predictive uncertainty. These uncertainties can then further be linked to a physical interpretation. As an example, the spatial UQ is shown in Figure 22, displaying the one-sided width of the 80% confidence interval (CI) of pressure and temperature (a derivation of this interval is found in [23]). The more uncertain areas in the temperature distribution correspond to the most dynamic sections of the transition region of the flowfield, where surface wave propagation is dominant. The downstream uncertainty is due to the recirculation caused by vortex breakdown. The pressure oscillations are more pronounced upstream, as shown through the UQ of the pressure distribution.

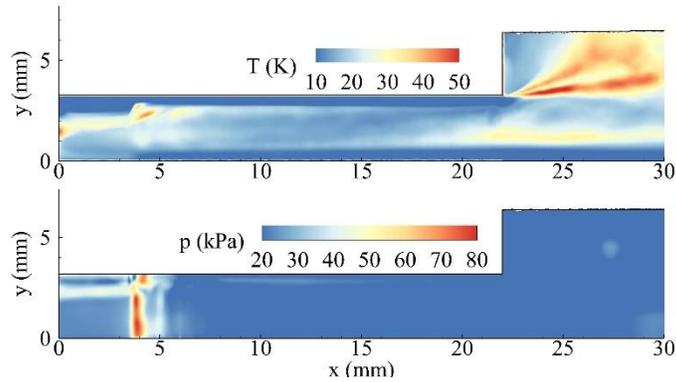

Figure 22. One-sided width of the 80% confidence interval for benchmark E: temperature and pressure predictions.

*3.8. Computation Time*

Figure 23 presents the simulation and emulation timeline. The computation times are calculated based on performance on a parallelized system of 200 Intel Xeon E5-2603 1.80 GHz processors. A total of 900,000 CPU hours is required for the 30 GB dataset. CPOD extraction and parameter estimation for the model takes about 45 minutes.

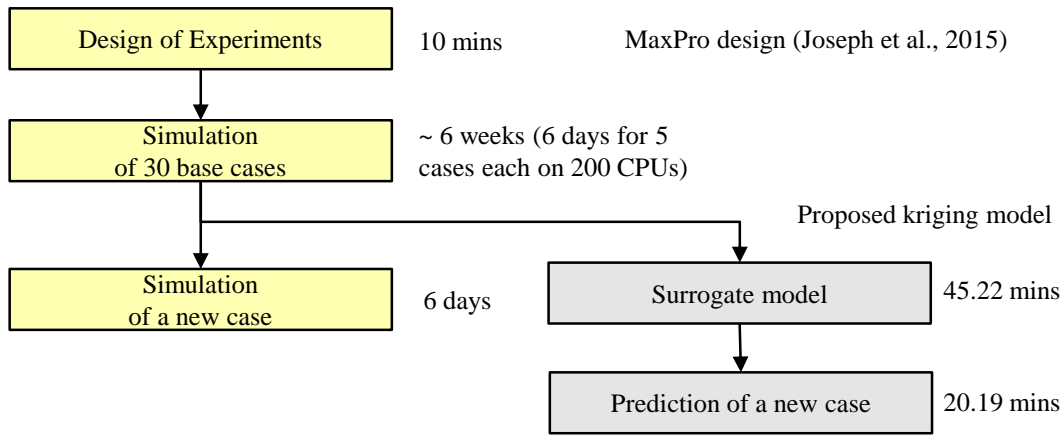

Figure 23. Simulation and emulation timeline.

The parallelized flowfield predictions from the proposed model require around 30 CPU hours, significantly reducing the turn-around time as compared with LES simulations requiring 30,000 CPU hours. This improved computational efficiency is crucial, as it enables quick turn-around times for design iterations. In comparison, the existing spatio-temporal emulators mentioned in the introduction require much more computation time to fit the underlying statistical model, because the training dataset of each simulation is too large to directly manipulate [17]. By carefully combining physical knowledge of the system with state-of-the-art machine learning techniques to make informed model assumptions, the proposed methodology offers an efficient strategy to survey the design space without the need for more expensive high-fidelity simulations.

## 4. Conclusion

The present work develops an integrated framework that incorporates state-of-the-art statistical methods, machine learning algorithms, and a physics-driven data reduction method to obtain a surrogate model over a broad range of design space. Taking a swirl injector as an example, the CPOD-based emulation framework is used to extract the flow physics, reduce the data, and build an efficient, physics-driven emulator.

The key contributions are two-fold: the use of statistical and machine learning techniques to quantify the impact of design parameters on important flow physics, and the incorporation of such methods with physics-guided model assumptions to build an efficient surrogate model for flowfield prediction. A vital model assumption is that the CPOD, the common basis, accurately retains the rich set of physics over varying geometries. This model successfully captured the simulation results and fared better than analytical estimations of the performance measures. The emulated flowfield is validated against the LES-simulated flowfield to demonstrate how the flow structures and injector characteristics are captured by the model. Moreover, this methodology significantly reduces the computational time required for surveying the design space for flow predictions. While the focus of the present paper is on a data-driven analysis and emulation of a swirl injector, the principle of applying machine learning techniques with physics-guided assumptions can be applied to any type of system model.

For future work, additional investigation should be carried out in dynamic regions of the flowfield, where the surrogate model had higher predictive uncertainties. One potential cause is the extreme range of the design points; this can be addressed by setting a smaller range. The uncertainty quantification and propagation of underlying flow couplings [47] are also important research directions, and can perhaps be tackled using techniques such as support points [48]. Another interesting path would be to improve the MaxPro methodology with minimax coverage [49]. Although the smoothing effect from CPOD-based emulation appears to be mitigated with proper parameter tuning, the extension of the decision tree to automatically determine the optimal basis functions for a prediction, such as hierarchical clustering [50], could significantly improve the methodology's capability of surveying large design spaces. From an implementation point of view, another hurdle is computational efficiency, where the use of local Gaussian Process models [51] appear to be an attractive option.

# Acknowledgement

This work was sponsored partly by the William R. T. Oakes Endowment and partly by the Coca-Cola Endowment of the Georgia Institute of Technology.